\documentclass[journal]{IEEEtran}

\ifCLASSINFOpdf
\else

\fi

\let\chapter\section
\usepackage{amsmath}
\usepackage{graphicx}
\usepackage{subfigure}
\usepackage{multirow}
\usepackage{amssymb}
\usepackage{algorithmic}
\usepackage{caption}
\usepackage{color}
\usepackage[ruled,lined]{algorithm2e}
\usepackage{bm}
\usepackage{enumerate}
\usepackage{mathrsfs}
\usepackage{leftidx}
\usepackage{authblk}
\usepackage{booktabs}
\usepackage{threeparttable}
\usepackage{array}
\usepackage{amsfonts,amssymb}
\usepackage{hyperref}
\usepackage{textcomp}
\usepackage{cite}

\usepackage{tikz}
\usetikzlibrary{positioning}
\newcommand{\tabincell}[2]{\begin{tabular}{@{}#1@{}}#2\end{tabular}} % 单元格 强制换行语句b

\begin{document}

\captionsetup[table]{name={TABLE},labelsep=none} %% reset TABLE.x: to TABLE x   labelfont={bf}加粗
\captionsetup[figure]{name={Fig.},labelsep=period} %% reset Fig.x: to Fig. x.

\title{Fast Convergence Time Synchronization in Wireless Sensor Networks Based on Average Consensus}

\author{

        Fanrong~Shi,~\IEEEmembership{Student Member,~IEEE},
        Xianguo~Tuo,
        Lili~Ran ,
        Zhenwen~Ren,~\IEEEmembership{Student Member,~IEEE},
        and~ Simon~X.~Yang,~\IEEEmembership{Senior Member,~IEEE}

\thanks{Manuscript received March 18, 2019; revised July 23, 2019; accepted August 14, 2019. This work was supported in part by the National Natural Science Foundation of China Programs under Grant 61601383, in part by the Sichuan Science and Technology Program under Grant 2018GZ0095, and in part by the Longshan academic talent research supporting program of Southwest University of Science and Technology under Grant 17LZX650 and Grant 18LZX633. Paper no. TII-19-0934. (Corresponding author: Xianguo Tuo.)}

\thanks{Fanrong Shi, Lili Ran, and Zhenwen Ren are with School of Information Engineering, University of Science and Technology, Mianyang 621010, China (e-mail: sfr\_swust@swust.edu.cn; lili\_ran@126.com; rzw@swust.edu.cn.}

\thanks{Xianguo Tuo is with Sichuan University of Science and Engineering, Zigong 643000, China (e-mail:tuoxg@cdut.edu.cn).}

\thanks{Simon X. Yang is with Advanced Robotics and Intelligent Systems (ARIS) Laboratory, School of Engineering, University of Guelph, Guelph, Ontario, N1G 2W1, Canada (email: syang@uoguelph.ca).}

}

\markboth{}  %Journal of \LaTeX\ Class Files,~Vol.~14, No.~8, August~2015
{Shell \MakeLowercase{\textit{et al.}}: Bare Demo of IEEEtran.cls for IEEE Journals}

\maketitle

\begin{abstract}Average consensus theory is intensely popular for building time synchronization in wireless sensor network (WSN). However, the average consensus-based time synchronization algorithm is based on iteration that pose challenges for efficiency, as they entail high communication cost and long convergence time in large-scale WSN. Based on the suggestion that the greater the algebraic connectivity leads to the faster the convergence, a novel multi-hop average consensus time synchronization (MACTS) is developed with innovative implementation in this paper.
By employing multi-hop communication model, it shows that virtual communication links among multi-hop nodes are generated and algebraic connectivity of network increases. Meanwhile, a multi-hop controller is developed to balance the convergence time, accuracy and communication complexity. Moreover, the accurate relative clock offset estimation is yielded by delay compensation. Implementing the MACTS based on the popular one-way broadcast model and taking multi-hop over short distances, we achieve hundreds of times the MACTS convergence rate compared to ATS.
\end{abstract}


\IEEEpeerreviewmaketitle

\section{Introduction}
\IEEEPARstart{S}ynchronization is an important part of the distributed system, e.g., synchronization control \cite{TIE_2}, synchronization measurement \cite{TIE_8}, internet of things \cite{TS-IOT,TII_6}. For the wireless sensor networks (WSN), time synchronization is used to generate consistent time concept for data acquisition \cite{TII_1,DataAqu,TII_2}, intelligent sleeping \cite{2}, location services \cite{3,Localtion_TII} and so on. Because of the extra hardware required or the complex protocols employing in the Network Time Protocol, Global Navigation Satellite System and IEEE 1588 Standard (Precision Time Protocol) may not meet the synchronization requirements in WSN applications.

In low-cost, low-power, energy-constrained, and large-scale WSN applications, the convergence rate is intensely important to an accurate, robust and energy-efficient time synchronization algorithm. A faster-convergence approach requires less synchronization intervals and message transmissions to obtain the expected synchronization accuracy, meanwhile, it may adapts quickly to the changes in network structure and clock drifts (e.g., due to variances in temperature or voltage).

Until now, numerous time synchronization algorithms have been proposed for WSN, and they can be simply summarized as centralized, semi-distributed, and fully distributed. The centralized time synchronization algorithms, e.g., the TPSN \cite{TPSN}, RBS \cite{RBS}, CESP \cite{CESP}, R-Sync \cite{R-Sync}, and cluster-based consensus time synchronization \cite{cluster-Conse-sync-1,cluster-Conse-sync-2,cluster-Conse-sync-3,PulseSS}, require additional topology management protocol (e.g., spanning tree protocol, and clustering protocol) to guarantee its validity. While, considering the large-scale WSN, the changes in the topology will cause the algorithms to fail. Thus, the centralized algorithms may well meet the time synchronization of lightweight static rather than large-scale dynamic WSN.

Both the semi-distributed and fully distributed time synchronization algorithms require none topology management protocol. The flooding time synchronization algorithms, e.g., FTSP \cite{FTSP}, EGSync \cite{EGSync}, Glossy \cite{Glossy}, PulseSync \cite{PulseSync}, FCSA \cite{FCSA}, PISync \cite{PISync}, and RMTS \cite{RMTS}, are the typical semi-distributed approach. The key idea is: a special node (i.e., reference) periodically broadcasts its time information, which is then flooded in network-wide, meanwhile other nodes try to synchronize themselves to the reference. While, in the fully distributed approaches, e.g., DCTS \cite{DCTS}, ATS \cite{ATS}, MTS \cite{MTS}, GTSP \cite{GTSP}, CCS \cite{CCS}, and DiStiNCT \cite{DiStiNCT}, none reference is required and each node synchronize to its neighbors, in other words, they use local information to achieve a global agreement. Therefore, the distributed approaches are more scalable and robust, and thus are paid more attention in the recent years.

The average consensus theory is widely used in many fields, e.g., distributed control \cite{cons-control-1,cons-control-4}, tracking \cite{consensus-tracking}, and smart grid \cite{SmartGrid_EnergyMan}. Moreover, it is popular used to develop a fully distributed time synchronization protocol for a dynamic WSN, e.g., the ATS. The average consensus-based time synchronization algorithm is easily to implement, but is more robust and scalable in dynamic networks than most of other time synchronization protocols. However, it is difficult to predict the convergence time  due to iteration, and needs to be improved in terms of convergence rate and synchronization accuracy.

Regarding synchronization convergence rate, as described in \cite{ATS}, in a $5\times7$ grid network (diameter of 10), ATS converges after 120 rounds of synchronization intervals. As shown in the experimental results of this paper, e.g., a $5\times5$ grid network (diameter of 8), the ATS requires approximately 24 rounds of synchronization intervals to convergent. While, considering the line network (diameter of 8), it requires about approximately 60 rounds of synchronization intervals. The maximum-consensus-based protocols MTS \cite{MTS} converge faster than ATS, but they still cost approximately 90 rounds to converge in a 20-node ring network (diameter of 10), and their convergence time is linear growth of the diameter. However, considering the flooding time synchronization algorithm, the FTSP and PulseSync can converge with less than 20 rounds of synchronization intervals in a 31-node line network (diameter of 29) \cite{FCSA}, and the RMTS can converge after the 2-nd synchronization interval in a 25-node line network (diameter of 24) \cite{RMTS}.

In terms of synchronization accuracy, ATS uses one-way broadcast model to generate the relative clock offset estimation; that is, the timestamps that are created by the broadcast frame at the transmitting node and the receiving node, respectively, are used to calculate the relative clock offset estimation. Unfortunately, in which the message delay is directly introduced into the estimation. In one-way broadcast time synchronization, there is currently no better way to eliminate the synchronization error that caused by delay, but uses previous experimental test results to estimate the mean of delay and then compensate the relative clock offset estimation by the mean, just like PulseSync and RMTS.

It has got detailed proof in \cite{14} that, the average consensus convergence rate equals to the algebraic connectivity of a strongly connected digraph. Moreover, the algebraic connectivity strongly depends on the structure and connections of the graph, and it is relatively large for dense graphs with more connections. Therefore, increasing the algebraic connectivity should be an efficient way to improve the convergence rate. In \cite{Conv-Analysis-Consensus}, it is recommended to increase the nearest neighbors or transmission radius without effecting the power consumption. While, considering the low-cost and energy-constrained WSN, it may be unreasonable due to the strong relationship between transmission distance and transmission power. In other words, physically changing the topology is difficult in WSN. The \cite{15} developed an multi-hop relay protocol for multi-agent system to get an better convergence rate without changing topology, in which a large number of virtual connection are generated to the improve on algebraic connectivity. However, considering the large-scale WSN, the multi-hop communication may seriously increases the message and computation complexity \cite{15}. Moreover, it may results in a by-hop error accumulation problem \cite{RMTS}.

In this paper, we focus on an accurate average consensus time synchronization algorithm that converges fast. The multiple-hop average consensus time synchronization (MACTS) algorithm is proposed based on multiple-hop one-way broadcast model, and the main important improvements of MACTS are as follows.

1) The convergence rate is greatly improved in MACTS. Multi-hop communication is used to generate virtual connection between multi-hop nodes, which may brings order of magnitude improvement in algebraic connectivity and convergence rate.

2) A multi-hop controller is developed to optimize drawbacks due to multi-hop communication in MACTS. To obtain a fast convergence, it employs a multi-hop communication at the initial phase of network; to simplify the message complexity and against the by-hop error accumulation, it adaptively switches to single-hop communication after the synchronization converges.

Moreover, a delay estimate is calculated from the results of previous experimental results, and it is used to compensate the relative clock offset estimation in MACTS.

The remainder of this paper is organized as follows. The system model is provided In Section II. We present and analyze the MACTS in Section III. The implementation of MACTS is developed in Section IV. The experimental results and simulation are presented and discussed in Section V and VI, respectively. Conclusions are given in Section VII.

\section{System Model}
The WSN is modeled as the graph $\mathcal{G}=(\mathcal{N},\mathcal{E})$, where $\mathcal{N}=\{1,2,\ldots,\mathcal{N}\}$ represents the nodes of the WSN and $\mathcal{E}$ defines the available communication links. The set of neighbors for $v_i$ is $\mathcal{N}_i=\{j\mid(i,j)\in \mathcal{E},i\neq j\}$, where nodes $v_i$ and $v_j$ belong to $\mathcal{N}$, and $j\in \mathcal{N}_i$.

There are two time notions defined for the time synchronization algorithm, i.e. the hardware clock notion $H(t)$, and logical clock notion $L(t)$. The $H(t)$ is defined as
\begin{equation}\label{equ:1}
 H(t)=\int_0^t h(\tau)d\tau+\theta(t_0),
\end{equation}
where $h(\tau)$ is the hardware frequency rate (clock speed) of the clock source. The $h(\tau)$ is the inherent attribute of crystal oscillator; it cannot be changed or measured. Any node considers itself having the ideal clock frequency (i.e., nominal frequency), and $h(\tau)=1$. The variable $t_0$ is the moment that node is powered on and $\theta(t_0)$ the initial relative clock offset. It should be noted that $H(t)$ cannot be changed also, and the timestamps on $H(t)$ are used to estimate the relative clock speed for the proposed algorithm.

The $L(t)$ is defined as
\begin{equation}\label{equ:2}
  L(t)=\varphi(t)\times H(t),
\end{equation}
where $\varphi(t)$ is the logical relative clock rate and initialized as 1; it can be changed to speed up or slow down $L(t)$. With respect to the reference node, $\varphi(t)$ is always set as 1. The timestamps on $L(t)$ are used to estimate the relative clock offset for the proposed algorithm. $L(t)$ is used to supply the global time service for the synchronization applications in WSN.

Considering the arbitrary nodes $v_i$ and $v_j$, $L_i(t)$ and $L_j(t)$ are the logical times, respectively, and $\varphi_i^j$ is the relative frequency rate, which is
\begin{equation}\label{equ:3}
\varphi_i^j=\frac{H_j(t_2)-H_j(t_1)}{H_i(t_2)-H_i(t_1)},		
\end{equation}
where, in consideration of arbitrary moments $t_1<t_2$.

\section{ The Proposed MACTS}
\subsection{The MACTS Protocol}
The key idea of the proposed MACTS is to increase the connection $\mathcal{E}$ by Building virtual communication links among multi-hop nodes, then increases the algebraic connectivity indirectly. While the added virtual connections dose not change the topology structure.

\begin{figure}[!htb]
\centering
\includegraphics[scale=1]{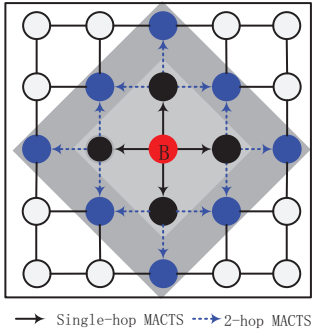}
\caption{The 2-hop MACTS indication ($H=2$). Considering the moment that $v_R$ broadcast a packet, its neighbor will forward the received packet to the multi-hop node.}
\label{fig.1}
\end{figure}

In MACTS, every node asynchronously and periodically broadcasts time synchronization messages to neighbor nodes. Considering a $5\times5$ grid networks, taking 2-hop MACTS as an example, it is shown in Fig. \ref{fig.1}. The solid arrows in the figure are 1-hop MACTS and the dotted lines show the 2-hops MACTS. The broadcast information of node $v_R$ is received and processed by its neighbor nodes (black nodes).

In mult-hops MACTS, neighbor nodes immediately forward the time information they received, as shown by the dotted arrows in the Fig. \ref{fig.1}. At this time, the synchronous broadcast message of $v_R$ is not only processed by its neighbor nodes, but also by the 2-hop nodes (blue nodes). MACTS builds virtual connections between node $v_R$ and 2-hop nodes to increase the number of network connections. Here, parameter $H$ is defined as the multi-hop number.

\vspace{0.3cm}

\begin{figure}[!htb]
\centering
\includegraphics[scale=0.6]{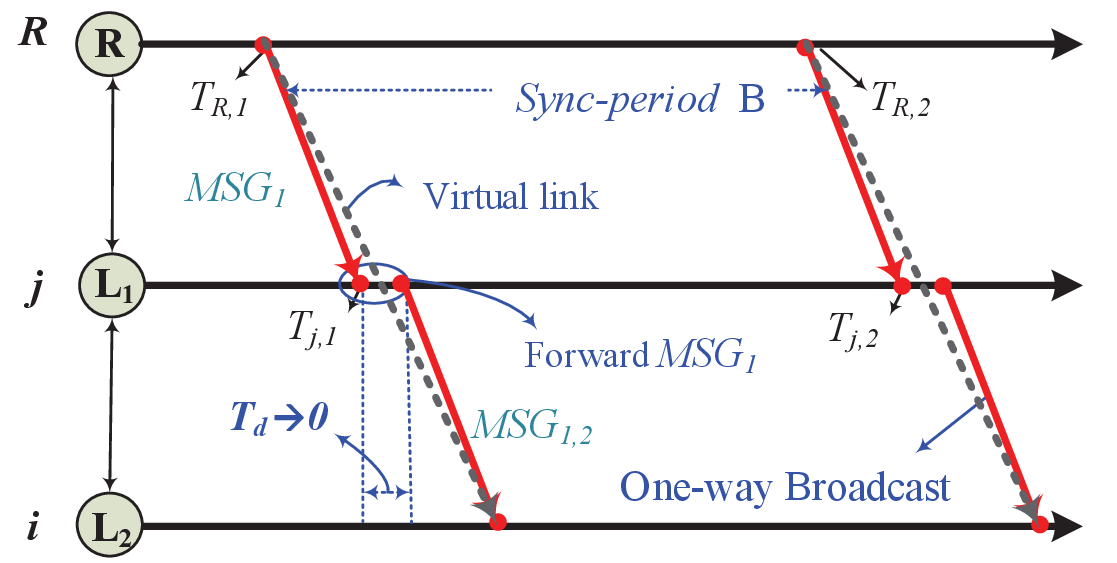}
\caption{The 2-hop MACTS synchronization model. The 1-hop node forwards the packet as soon as possible, i.e., $T_d\rightarrow0$.}
\label{fig.2}
\end{figure}

Figure \ref{fig.2} shows the implementation of multi-hop communication in MACTS. When the synchronization cycle of $v_R$ is triggered, its broadcast time message frame, i.e., $MSG_1$, will generate time stamps $T_{R,1}$ and $T_{j,1}$ in $v_R$ and $v_j$, respectively. In multi-hop MACTS, $v_j$ will forward the $MSG_1$ as soon as possible, i.e., the latency time $T_d\rightarrow0$, then $MSG_{1,2}$ is forwarded to $v_i$. Thus, there is a virtual communication link is generated between $v_R$ and $v_i$.

\vspace{0.3cm}

The time information of the broadcast node is forwarded to the multiple-hop nodes by adjacent nodes, since there is many virtual communication links between broadcast and multiple-hop nodes; and the algebraic connectivity will be increased greatly.

\textbf{\emph{1) Relative clock skew estimation:}} as shown in Fig. \ref{fig.2}, according (\ref{equ:3}), the relative frequency rate estimation $\hat{\varphi}_R^j$ is given by
\begin{equation}\label{equ:4}
\hat{\varphi}_j^R=\frac{T_{j,2}-T_{j,1}}{T_{R,2}-T_{R,1}},
\end{equation}
where the timestamps are created by the hardware clock $H(t)$.

\textbf{\emph{2) Relative clock offset estimation:}} the clock offset estimation $\hat{\theta_j^R}$ is usually given by $\hat{\theta_j^R}=T_{j,1}-T_{R,1}$, where $T_{j,1}-T_{R,1}=\theta_j^R+D$. Variable $D$ is defined as the difference of a pair of time stamps, and the $D$ is comprised of the fixed portion $D_{fixed}$ and the variable portion $d$ \cite{12,13}, i.e., $D=D_{fixed}+d$. In our experimental results, $\bar{D}$ is about 3.33 {\textmu s}, $\sigma$ is 0.07, and $D$ is the normal distribution with 99.99\% confidence determined by a t-test. Setting the fixed portion delay estimation $\hat{D}_{fixed}=\bar{D}$, then the $\hat{\theta_j^R}$ is
\begin{equation}\label{equ:5}
 \hat{\theta}_j^R=T_{j,1}-T_{R,1}-\bar{D}=\theta_R^j+(D-\hat{D}_{fixed}),
\end{equation}
where the timestamps are created by the logical clock $L(t)$. It can be seen that the estimation error of $\hat{\theta}_R^j$ is about $d$. Therefore, after $\hat{\varphi}_j^R$ converges, the MACTS can obtain a higher-precision relative clock offset estimation.

\textbf{\emph{3) Synchronization error analysis:}}
Base on the clock skew estimation and clock offset estimation, the local synchronization error in MACTS is given by
\begin{equation}\label{equ:45}
 e_l(t)=(D-\hat{D}_{fixed})+({\varphi}_j^R-\hat{\varphi}_j^R)\times t,
\end{equation}
where, $t$ is the time after latest synchronization, i.e., $0<t<B$.

According the analysis in \cite{RMTS}, the by-hop error accumulation in the $k$-hop MACTS is given by
\begin{equation}\label{equ:45}
  e_m[k]=\sum_{h=0}^{k-1}(D[h]-\hat{D}_{fixed})+\frac{\sum_{h=0}^{k-1}{(|a[h]-\hat{a}[h]|\times T_d)}}{10^6},	
\end{equation}
where, $T_d \rightarrow0$ is the waiting time of multi-hop node forward the time information.

\subsection{Averaging Problem and Convergence}
\textbf{\emph{1) Average-consensus:}}
in \cite{11,14}, the averaging problem is discussed in detail. Defining $x_i(0)$ as the initial state of node $i\in V$ and the state vector $x(t)=\{x_1(t),x_2(t),\ldots,x_n(t)\}$ of the node at time $k$, a consensus algorithm to update the state through the dynamic equation $x(t)$ is given as follow
\begin{equation}\label{equ:6}
    x(t+1)=A(t)x(t).
\end{equation}

The definition of matrix $A(t)$ is a non-negative matrix composed of $a_{ij}(t)$, which is a weighted adjacency matrix of $G$, and has $\sum_{j=1}^na_{ij}(t)=1, (n\leq \mathcal{N})$ for any $v_i$; that is, $A(t)$ is a stochastic matrix. Then, the dynamic equation of the discrete consensus principle of any node i of the WSN is as follows \cite{14}
\begin{equation}\label{equ:27}
    x_i(k+1)=x_i(k)+\sum_{j\in \mathcal{N}_i}a_{ij}(k)(x_j(k)-x_i(k)).
\end{equation}

For the static WSN $\mathcal{G}$, the set of its connection, $\mathcal{E}$, is a fixed value, and then the adjacency matrix is a fixed value; that is, $A(t)=A$. From \cite{11}, $\lim\limits_{{t\rightarrow\infty}}x_i (t)=\sum_{i=1}^n\pi_ix_i(0)$, where $\pi_i$ is the steady-state probability of the $v_i$ in the Markov chain associated with the stochastic matrix $A$. Then, if and only if any $v_i$ has $\pi_i=1/n$, the average consensus convergence is obtained. Therefore, using a doubly stochastic matrix, all nodes $v_i$ can converge to
\begin{equation}\label{equ:8}
    \lim_{t\rightarrow\infty}x_i(t)=\frac{1}{n}\sum_{i=1}^nx_i(0),
\end{equation}
where every node converges to the average of the initial state of all nodes. Specific proofs can be seen in \cite{11,14}.

\textbf{\emph{2) Algebraic connectivity and convergence rate:}}
The convergence rate and convergence time of the average consensus algorithm directly depend on the algebraic connectivity of the $\mathcal{G}$. That is \cite{14},
\newtheorem{lemma}{\textbf{Lemma}}[section]
\begin{lemma} \label{lemma1}
The more network connections and the greater the algebraic connectivity, the faster the convergence rate.
\end{lemma}

It defines that, a network with a fixed topology that is a strongly connected digraph, the globally asymptotically vanishes with a speed equal to the algebraic connectivity. The algebraic connectivity is determined by the topology and the edges of graphs, and it is relatively large for dense graphs (more edges), and is relatively small for sparse graphs (less edges) \cite{14}.

Considering node $v_i$ in $\mathcal{G}$, its size of neighbors is $|\mathcal{N}_i|$, and the in-degree and out-degree of node $v_i$ are
\begin{equation}\label{equ:100}
   deg_{in}(v_i)=\sum_{j=1}^na_{ji}, deg_{out}(v_i)=\sum_{j=1}^na_{ij}.
\end{equation}
\\Because $\mathcal{G}$ is symmetrically connected, there are node connections that are bidirectional; that is, both the in-degree and the out-degree are equal to $|\mathcal{N}_i|$. The degree matrix $\mathcal{D}$ is a diagonal matrix. Then the Laplacian matrix $\mathcal{L}$ of the $\mathcal{G}$ is composed of the $A$ and $D$ of $\mathcal{G}$; that is
\begin{equation}\label{equ:11}
    \mathcal{L}=\mathcal{D}-A.
\end{equation}

The graph $\mathcal{G}$ has a weighted adjacency matrix $a_{ij}=a_{ij}\cdot w_{ij}$, where $w_{ij}>0$ is the associated weight of the pair of nodes $(i,j)$. The sum of arbitrary rows of the Laplacian matrix $\mathcal{L}$ is 0, and is a semi-positive definite matrix with $n$ non-negative eigenvalues $0=\lambda_1\leq\lambda_2\leq \ldots\leq\lambda_n$, where the second-smallest eigenvalue $\lambda_2$ is the algebraic connectivity of $\mathcal{G}$.

According to \cite{15}, for the symmetric directed graph $\mathcal{G}$, $\lambda_2$ can obtain the maximum value when all weights are the same; that is,
\begin{equation}\label{equ:10}
   \lambda_2\leq\sum_{i\neq j}w_{ij}/(n-1).
\end{equation}
Thus, when the directed graph $\mathcal{G}$ is set to the same weight $w_{ij}$, the average consensus protocol can obtain the maximum convergence rate. In the MACTS $w_{ij}$ is set to the constant 1.

According \cite{14,15}, the algebraic connectivity $\lambda_2$ is determined by the topology structure of $\mathcal{G}$ and the connection $\mathcal{E}$, moreover, the $\lambda_2$ of a dense graph is relatively lager than that of a sparse graph.

\subsection{The Convergence Rate Analysis}
The \cite{Conv-Analysis-Consensus} suggests that, the number of neighbors and connections could be increased by increasing transmit power. However, in a static wireless sensor network, the network topology is fixed, and the transmit power (distance) of node is limited. In a dynamic wireless sensor network, the network topology dynamically changes randomly, and it cannot change according to the trend that is beneficial to increase the algebraic connectivity. Therefore, changing network topology is not the best choice to improve the convergence rate and does not meet the actual applications requirements.

In the proposed MACTS, the status information of the broadcast node is received not only by its neighbor nodes, but also by the multi-hop node. These multi-hop nodes constitutes virtual connection $\mathcal{E}$. It can be derived that the multi-hop consensus dynamic equation is \cite{15}
\begin{equation}\label{equ:9}
     \begin{aligned}
 x_i(k+1)=x_i(k) + &\sum_{j\in \mathcal{N}_i}a_{ij}(k)(x_j(k)-x_i(k)+ \\
                   &  \sum_{s\in \mathcal{N}_j}a_{js}(k)(x_s(k)-x_j(k)+\ldots)).
    \end{aligned}
\end{equation}

For the directed graph $\mathcal{G}$, the directed graph is $\mathcal{G}_1$ when $H=1$, the directed graph is $\mathcal{G}_2$ when $H=2$, the directed graph is $\mathcal{G}_3$ when $H=3$, and so on. It can be seen from the above dynamic equations that the start and end points are links of the same node, which is a self-loop, and do not contribute to convergence, so these links are ignored in the subsequent discussion. In a multi-hop graph, there may be multiple H-hop links for pair of points, which are treated as paths here, and its weight is the sum of weights of all $H$-hop links. The algebraic connectivity of $\mathcal{G}_1$ is $\lambda_2(1)$.

Therefore, the adjacency matrix $A_2$ of $\mathcal{G}_2$ is  \cite{15}
\begin{equation}\label{equ:12}
    a_{is}=\left\{
    \begin{aligned}
    &\sum_{j\in N_i}a_{ij}a_{js} ,  (v_i,v_s)\in \mathcal{G}_2 \\
    &0
    \end{aligned}
    \right.
\end{equation}

The corresponding Laplacian matrix is $\mathcal{L}_2$, the degree matrix is $D_2$, and the corresponding algebraic connectivity is $\lambda_2(2)$.

Similarly, the adjacency matrix $A_3$ of $\mathcal{G}_3$ is
\begin{equation}\label{equ:13}
    a_{iz}=\left\{
    \begin{aligned}
    &\sum_{j\in N_j}a_{ij}(\sum_{s\in N_j}a_{js}a_{zs}),  (v_i,v_z)\in \mathcal{G}_3 \\
    &0
    \end{aligned}
    \right.
\end{equation}
The corresponding Laplacian matrix is $\mathcal{L}_3$, the degree matrix is $D_3$, and the corresponding algebraic connectivity of $\mathcal{G}_H$  is $\lambda_2(H)$.

Let the column vector $x$ of the matrix $W$ satisfy $xx^T=1$, and the sum of $x$ is 0. Based on the Courant-Fischer theorem, the second-smallest eigenvalue $\lambda_2$ of Laplacian matrix $\mathcal{L}$ is given by the following equation
\begin{equation}\label{equ:14}
    \lambda_2=\min\limits_{x\in W}x^TLx.
\end{equation}

It is concluded that if $\mathcal{G}_1$ and $\mathcal{G}_2$ have the same vertex, $\mathcal{G}=\mathcal{G}_1\cup \mathcal{G}_2$, and $a(\mathcal{G})$ is the solution of the eigenvalue $\lambda_2$; then \cite{15},
\begin{equation}\label{equ:15}
    a(\mathcal{G}_1)+a(\mathcal{G}_2)\leq a(\mathcal{G}_1\cup \mathcal{G}_2).
\end{equation}
The above conclusions can be generalized: If \{$\mathcal{G}_1$,$\mathcal{G}_2$,$\ldots$,$\mathcal{G}_H$\} satisfy the above conditions, then
\begin{equation}\label{equ:16}
    a(\mathcal{G}_1)+a(\mathcal{G}_2 )+\ldots+a(\mathcal{G}_H )\leq a(\mathcal{G}_1\cup \mathcal{G}_2\cup \ldots \cup \mathcal{G}_H ).
\end{equation}
The similar conclusions are proved in \cite{15}.

Therefore, the algebraic connectivity $\lambda_2$ of the H-hop consensus described above satisfies the following formula
\begin{equation}\label{equ:17}
    \lambda_2\geq\lambda_2(1)+\lambda_2(2)+\ldots\lambda_2(H).
\end{equation}

As the hop $H$ increases, the algebraic connectivity $\lambda_2$ of $\mathcal{G}$ gradually increases, thereby improving the convergence rate of the consensus time synchronization.

\section{Implementation}

\subsection{The Multi-hop MACTS}
The pseudo-code of the MACTS is presented in Algorithm 1. Node $v_i$ is synchronized to the $v_j$ by calibrating $L_i(t)$ based on the rate multiplier $\varphi_i$ and clock offset estimation $\hat{\theta}_i^j$, where $v_j$ is the neighbor of $v_i$.

\emph{\textbf{The multi-hop communication}}: node $v_i$ is broadcast periodically (period of $B$) to distribute the time information packets to neighbors, as in Algorithm 1, Line \ref{alg1.3}. Once the broadcast task is triggered, $v_i$ rapidly broadcasts a packet, as in Algorithm 1, Lines \ref{alg1.4} and \ref{alg1.5}. Two group of timestamps are created over the phase of broadcasting, i.e., timestamp $H_i[n]$ (created on $H_i(t)$) and timestamp $L_i[n]$ (created on $L_i(t)$). The basic information of the broadcast packets comprise four parts: $H_i[n]$, $L_i[n]$, $\varphi_i$, and $ID_{scr}$ (Algorithm 1, Line \ref{alg1.5}). The $ID_{scr}$ is used to count the forward distance (hop) of the time information, and if $ID_{scr}$ achieves to the $H$, then the current multi-hop communication will be stopped, as in Algorithm 1, Lines \ref{alg1.14}, \ref{alg1.12} and \ref{alg1.13} .

\LinesNumbered
\begin{algorithm}[ht]
\caption{Algorithm pseudo-code for MACTS (taking 2 hops MACTS as an example, i.e., $H=2$ ), where $v_j\in N_i$.}
\textbf{Initialization:}                                                        \\{\label{alg1.1}}
\quad{Set $\varphi_i=1$, $\hat{\varphi}_i^j=1$, $\hat{\theta}_i^j=0$, $H=2$}    \\{\label{alg1.2}}
\quad{Set $H_{i,old}=0$, $H_{j,old}=0$}                                         \\{\label{alg1.2}}
\quad{Start periodic broadcast task, period of $B$}                                \\{\label{alg1.3}}
\BlankLine
\if\textbf{ Upon triggering of broadcast task}{                                  \\{\label{alg1.4}}
\quad {Broadcast $\langle H_i[n],L_i[n],\varphi_i, ID_{Scr}\rangle $}           \\{\label{alg1.5}}
}
\BlankLine
\if\textbf{ Upon receiving $\langle H_j[n],L_j[n],\varphi_j, ID_{Scr}\rangle $}{         \\{\label{alg1.6}}
\quad       {Store $\langle H_i[n],H_j[n]\rangle$, $\langle L_i[n],L_j[n]\rangle$}     \\{\label{alg1.7}}
\quad       {Store $\varphi_j$ }                                                       \\{\label{alg1.8}}
\quad       {Calculate $\hat{\varphi}_i^j$, $\hat{\theta}_i^j$ }                      \\{\label{alg1.9}}
\quad       {Update $\varphi_i\leftarrow \langle \hat{\varphi}_i^j, \varphi_j\rangle$ }  \\{\label{alg1.10}}
\quad       {Update $ID_{Scr}\leftarrow ID_{Scr}+1$ }                                 \\{\label{alg1.14}}
\quad       {Update $H_{i,old}\leftarrow H_i[n]$, $H_{j,old}\leftarrow H_j[n]$}       \\{\label{alg1.11}}
   }
\If{\textbf{($ID_{Scr}<H$})}{                                                           {\label{alg1.12}}
             {Broadcast $\langle H_i[n],L_i[n],\varphi_i, ID_{Scr}\rangle $}            \\{\label{alg1.13}}
            }
\end{algorithm}

\emph{\textbf{Relative clock skew estimation}}: The rate multiplier $\varphi_i$ of the logical clock $L_i(t)$ is initialized as 1 (Algorithm 1, Line \ref{alg1.2}), and will be shared with the neighbors as a part of the time synchronization information packet (Algorithm 1, Lines \ref{alg1.5} and  \ref{alg1.13}). The relative clock skew estimation $\hat{\varphi}_i^j$ is initialed as 1 (Algorithm 1, Line \ref{alg1.2}), and calculated by (\ref{equ:4}) (Algorithm 1, Line \ref{alg1.9}). For the purpose that the $L_i(t)$ runs at the same rate as $L_j(t)$, $\varphi_i$ will be updated by multiplying $\hat{\varphi}_i^j$ and $\varphi_j$, as in Algorithm 1, Line \ref{alg1.10}. Here, the hardware clock timestamps are used to calculate $\hat{\varphi}_i^j$ , i.e., $\langle H_i[n],H_{i,old}\rangle$ and $\langle H_j[n],H_{j,old}\rangle_{old}$. The $\varphi_i$ is given by
\begin{equation}\label{equ:6}
    \varphi_i=\rho_v \times \varphi_i + (1-\rho_v)\times (\hat{\varphi}_i^j \times \varphi_j),
\end{equation}
where $\rho_v\in(0,1)$ is the average factor, and is set as 0.5 in MACTS that similar to ATS.

\emph{\textbf{Relative clock offset estimation:}} The parameter $\hat{\theta}_i$ is calculated by (\ref{equ:5}), as in Algorithm 1, Line \ref{alg1.9}. The logical timestamps $\langle L_i[n], L_j[n]\rangle$ are used, i.e., $\hat{\theta}_i^j=L_i[n]-L_j[n]-\hat{D}_{fixed}$, where the $\hat{D}_{fixed}$ is a constant that is calculated based on the precious experimental results.

\emph{\textbf{Logical clock maintenance}}: After $\varphi_i$ and $\hat{\theta}_{i(MLE)}^j$ are calculated and the local logical clock has been compensated, then the logical time will be updated by
\begin{equation}\label{equ:7}
 L_i(t)=(L_i(\tau)+\hat{\theta}_i^j/2)+(H_i(t)-H_i(\tau))\times\varphi_i, (t>\tau).
\end{equation}

\subsection{The MACTS Multi-hop Controller}
The MACTS expects to speed up the time synchronization convergence rate by employing multi-hop communication. However, when relay node forwards the multi-hop messages, the message complexity will inevitably increase, e.g., the increases in the number of message and the collision probability. Specifically, the more the number of the neighbor and multi-hop, then the more number of message, and more payload result in lager collision probability (e.g., refer to the transmission success probability model of Pure ALOHA protocol). Moreover, the multi-communication will cause to by-hop error accumulation along the multi-hop path \cite{RMTS}.

\begin{figure}[!htb]
\centering
\includegraphics[scale=1.1]{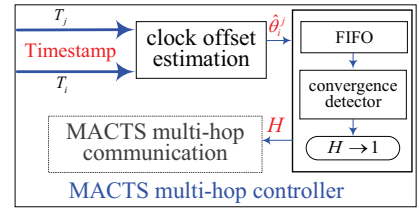}
\caption{The MACTS multi-hop controller.}
\label{multi-hop-controller}
\end{figure}

To overcome the above problems, a multi-hop controller is designed and demonstrated at Fig. \ref{multi-hop-controller}. In which, the key idea is: the MACTS runs on multi-hop algorithm at the initialization phase, so that it will converge fast; then it will switch to single-hop algorithm once the time synchronization convergent, and runs with low message complexity.

\vspace{0.5cm}

Variable $\hat{e}_i^j$ is the absolute relative clock offset estimation (synchronization error) among $v_i$ and $v_j (j\in N_i)$, and $\hat{e}_i^j$ is given by
\begin{equation}\label{equ:71}
 \hat{e}_i^j=|T_i-T_j|.
\end{equation}
The convergence detector is developed to detect the time synchronization convergence among $v_i$ and $v_j (j\in N_i)$. In other words, it is a local convergence detector. If $\{\hat{e}_i^j\}_{j\in N_i}<\xi$ is true, where parameter $\xi$ is threshold of local synchronization error (set as 5 {\textmu s} based on the previous experimental results), then detector will output a state of the local synchronization convergence, and multi-hop controller will reduces $H$ by 1 (if $H>1$); else, detector will output a state of the local synchronization failed, and multi-hop controller will increases $H$ by 1 (if $H$ is smaller than its initial value).

Moreover, suitable initial $H$ should be designed based on the estimate on the network size, topology, and diameter in advance. Specifically, the initial $H$ should be as small as possible in the large denser network; while should be as large as possible, so that the MACTS can converge faster.

\section{Experimental Results and Discussion}
Physical experiments were used to evaluate the synchronization accuracy and the convergence rate of the proposed MACTS. A prototype system was built around synchronous sensing wireless sensor (\emph{SSWS}) nodes. The \emph{SSWS} node is designed based on CC2530 with an enhanced 8051 CPU. An external 32-MHz crystal oscillator was set as the system clock source. The transmission power is programmed as 0 dBm. Timer 2 (with a 16-bit timer and a 24-bit overflow counter) of CC2530 works on up mode and was employed to generate the clock notion, and the clock granularity is 1 {\textmu s}.

\emph{\textbf{MAC-Layer timestamp:}} it is created by the Start Frame Delimiter (SFD) interrupt response function. The main uncertainly time of the timestamp is resulted from the interrupt handling time and response time. The difference of delay between sender and receiver present typical statistical characteristics. In our experiments, the $\hat{D}_{dixed}$ is about 3 {\textmu s} \cite{RMTS}.

\emph{\textbf{Synchronization error measurement:}} same to \cite{RMTS}, a sink-node that is connected to a PC is employed to measure the synchronization error. It periodically (interval of 10 seconds) transmits measurement command. All of the testbed nodes will generate a timestamp immediately when they receive the command, and then upload the timestamp to calculate the synchronization error. Experimental results in \cite{RMTS} indicate that, this measurement method is reliable and its measurement error has mean of 0.07 {\textmu s} and maximum of 0.4 {\textmu s}.

\begin{figure}[!htb]
\centering
\includegraphics[scale=0.5]{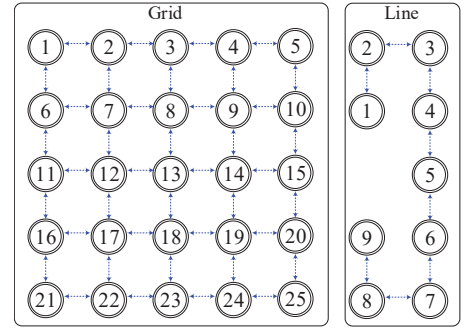}
\caption{The layout and links of the experimental testbed. The diameter is the same to the Grid and the Line}
\label{network}
\end{figure}

The layout and links of the indoor experimental testbeds are shown in Fig. \ref{network}. The $5\times5$ Grid (diameter of 8) testbed is made up of 25 \emph{SSWSs}, and it is employed to evaluate the performances of algorithms on a complex network (or dense network). The Line network are designed with 9 $SSWSs$, which has same diameter to the Grid, and it is employed to evaluate simple network (or sparse network) and the effects of the networks shape to the convergence rate and accuracy.

\subsection{Experimental Results on $5\times5$ Grid}
The experiments were performed for more than 260 minutes, and the synchronization period is 30 seconds.

\begin{figure}[!htb]
\centering
\includegraphics[scale=0.7]{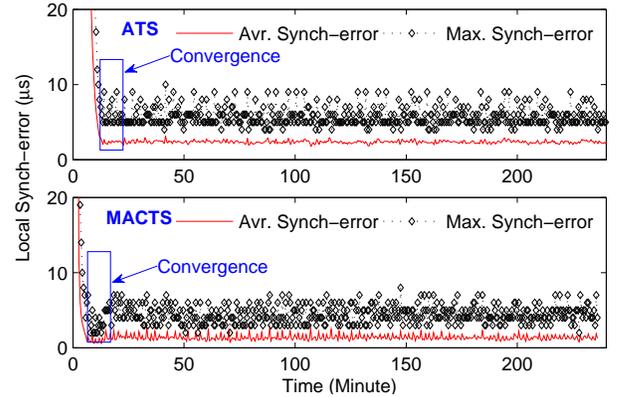}
\caption{Local clock synch-error on the $5\times5$ grid.}
\label{LocalSkew}
\end{figure}

\emph{\textbf{Local synchronization error}}: it is employed to evaluate the synchronization accuracy between any two adjacent nodes. The experimental results that are shown in Fig. \ref{LocalSkew} indicate the instantaneous average and instantaneous maximum of local synchronization error. The $D_{fixed}$ is introduced into the ATS's clock offset estimation $\hat{\theta}$ directly, while the MACTS uses the $\hat{D}_{fixed}$ to compensate $\hat{\theta}$ and the more accurate results are created for MACTS. After the algorithm enters a steady state, i.e., convergence, the maximal local synchronization error is about 10 {\textmu s} in ATS, and about 8 {\textmu s} in MACTS.

\begin{figure}[!htb]
\centering
\includegraphics[scale=0.6]{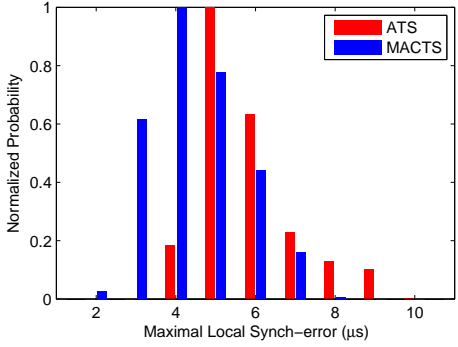}
\caption{The actual distributions of maximal local synch-error on the $5\times5$ grid.}
\label{LocalSkew_HIST}
\end{figure}

For the accurate time synchronization algorithm, the synchronization error is expected to be zero. The actual distributions of maximal local synchronization error are demonstrated in Fig. \ref{LocalSkew_HIST}, and the maximal distribution (probability) of the synchronization error is at 5 {\textmu s} in ATS, and at 4 {\textmu s} MACTS. Moreover, the probability distribution in MACTS is closer to zero, and its main part is smaller than 6 {\textmu s}. The 95\% confidence interval of the maximal local synchronization error is  mean 4.35-4.57 {\textmu s} and standard deviation 1.1-1.26 in MACTS, while it is, respectively,  5.69-5.9 {\textmu s} and standard deviation 1.26-1.41 in ATS. Therefore, the MACTS could yields better local synchronization than ATS.

\vspace{0.5cm}

\vspace{0.5cm}

\begin{figure}[!htb]
\centering
\includegraphics[scale=0.6]{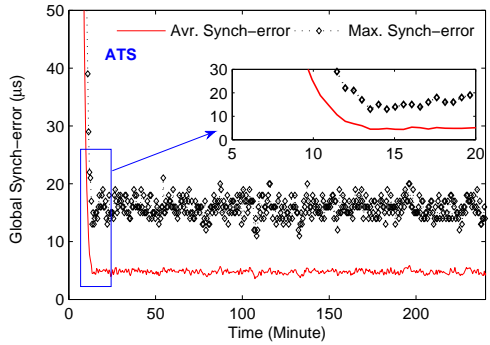}
\caption{Maximal global synch-error of ATS on the $5\times5$ grid. Zoom window shows the synchronization convergence time.}
\label{ATS_Global}
\end{figure}

\begin{figure}[!htb]
\centering
\includegraphics[scale=0.6]{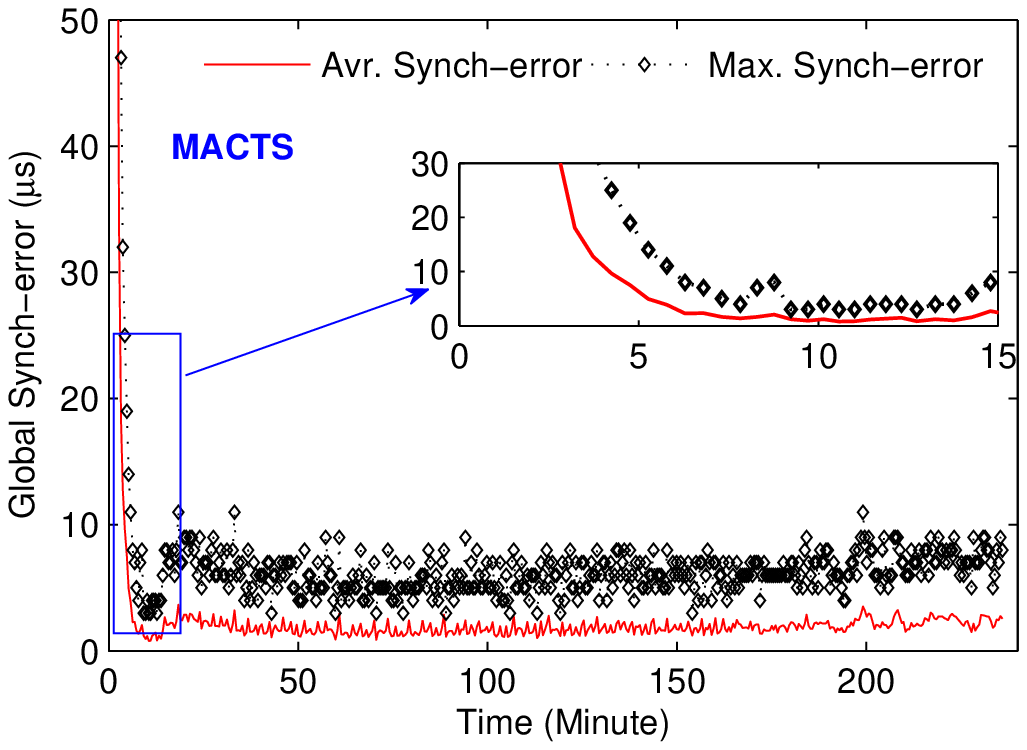}
\caption{Maximal global synch-error of MACTS on the $5\times5$ grid. Zoom window shows the synchronization convergence time.}
\label{MACTS_Global}
\end{figure}

\emph{\textbf{Global synchronization error}}: it is employed to evaluate the synchronization error between arbitrary pairs of nodes. The experimental results that are shown in Figures \ref{ATS_Global} and \ref{MACTS_Global} indicate the instantaneous average and instantaneous maximum of global synchronization error.

The maximal global synchronization error of ATS is less than 20 {\textmu s} in approximately 12 minutes. At this time, it is judged that ATS completes the time synchronization convergence; MACTS reaches the same synchronization error in approximately 5 minutes. In the dozens of experimental test results not reported in this paper, the convergence time is consistent with the above and there are a small differences because of the initial values of the relative time offset between nodes (node power-on time); the convergence time of ATS is approximately 10–14 minutes and the convergence time of MACTS is approximately 4–7 minutes.

\vspace{0.2cm}

\vspace{0.5cm}

\vspace{0.8cm}
\begin{table}[htbp]
 \centering
 \captionsetup{justification=centering}
 \caption{\\Summary of the maximal global synchronization error on the $5\times5$ grid. The 95\% confidence interval is mean 15.81-16.09 {\textmu s} and standard deviation 1.6-1.79 in ATS, mean 6.04-6.31 {\textmu s} and standard deviation 1.39-1.59 in MACTS.}{\label{tab:1}}
 \setlength{\tabcolsep}{0.5mm}{
 \begin{tabular}{ccccc}
  \toprule
                  &    \multicolumn{3}{c}{\textbf{{The maximal global Synch-error}}}                                     & \textbf{Convergence}\\
                  & \tabincell{c}{Mean (\textmu s)}    &\tabincell{c}{STD} &\tabincell{c}{Maximal (\textmu s)}          & \textbf{time (Minute)}\\
  \midrule
        ATS	      &15.95	                             &1.69	            &21	                                         & 10-14  \\
        MACTS	  &6.18	                                 &1.48	            &11	                                          &  4-7 \\
  \bottomrule
 \end{tabular}}
\end{table}
\vspace{0cm}

It can be seen that the convergence rate of MACTS is multiple-fold of ATS when $H=2$. Considering the synchronization accuracy after convergence, the Figures \ref{ATS_Global} and \ref{MACTS_Global} show that the maximal synchronization error and the average synchronization error of MACTS are better than those of ATS. The statistical characteristics of maximal global synchronization error are shown in Table \ref{tab:1}, from which one can see that the synchronization accuracy of MACTS is better than that of ATS.

\begin{figure}[!htb]
\centering
\includegraphics[scale=0.6]{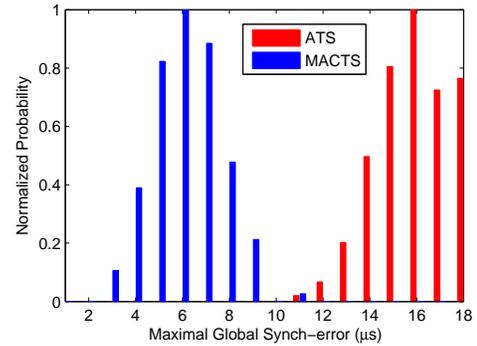}
\caption{The distributions of maximal global synch-error on a grid.}
\label{GlobalSkew_HIST}
\end{figure}

\begin{figure}[!htb]
\centering
\includegraphics[scale=0.6]{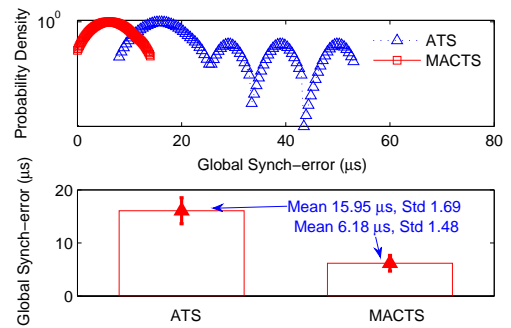}
\caption{The probability density and error bars of maximal global synch-error on grid. Logarithmic coordinates are used for Y-axis to show more details of probability density.}
\label{PDF_ErrorBAr}
\end{figure}

The actual distributions of maximal global Synchronization error are demonstrated in Fig. \ref{GlobalSkew_HIST}. The distribution is deviate from zero in ATS, while it is steeper and closer to zero in MACTS. It can be found that MACTS outperforms the ATS, and the value and probability of maximal global synchronization error in MACTS is smaller. The estimation of probability density distributions are plotted in upper panel of Fig. \ref{PDF_ErrorBAr} to show more statistical characteristics. It shows that probability density of maximal global synchronization error in MACTS is tighter, and the maximal global synchronization error is more concentrated around the mean. The error bar of the global synchronization error is plotted in lower panel of Fig. \ref{PDF_ErrorBAr}. The error bars also show the confidence intervals of the global synchronization error, and the proposed MACTS contributes the better accuracy than ATS.

\vspace{0.5cm}

\vspace{0.5cm}

\subsection{Experimental Results on Line}
The experiments were performed one Line for more than 240 minutes, and the synchronization period is 30 seconds.

\begin{figure}[!htb]
\centering
\includegraphics[scale=0.6]{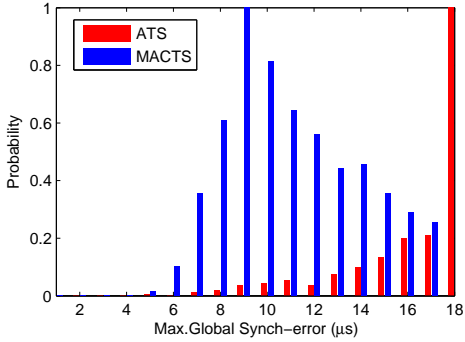}
\caption{The distributions of maximal global Synch-error on line.}
\label{Global_HIST_Line}
\end{figure}

\begin{figure}[!htb]
\centering
\includegraphics[scale=0.6]{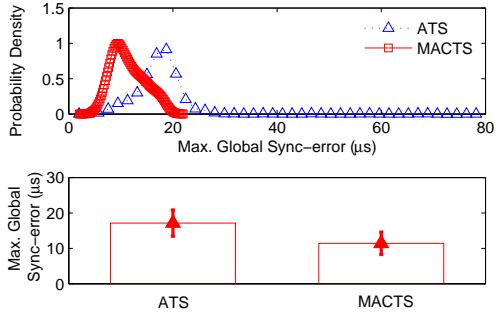}
\caption{The probability density and error bars of maximal global Synch-error on line.}
\label{PDF_ERROBAR_LINE}
\end{figure}

The actual distributions of maximal global Synchronization error are demonstrated in Fig. \ref{Global_HIST_Line}. The estimation of probability density distributions are plotted in upper panel of Fig. \ref{PDF_ERROBAR_LINE} to show more statistical characteristics. The MACTS's probability density is tighter and closer to zero. The error bar of the global synchronization error is plotted in lower panel of Fig. \ref{PDF_ERROBAR_LINE}, which shows that the confidence intervals of the global synchronization error. The statistical characteristics of maximal global synchronization error are shown in Table \ref{tab:2}, from which it can be found that the MACTS performances better than ATS in the line topology.

In conclusion, the different network structures (or shapes) will result in different performances for both MACTS and ATS, and the better accuracy and the faster convergence the algorithms are when used grid. Moreover, the MACTS show better performances than ATS in both grid and line. Thus, when the topology changes, MACTS can produce better performance.

\vspace{0.5cm}

\vspace{0.5cm}

\vspace{0.5cm}
\begin{table}[htbp]
 \centering
 \captionsetup{justification=centering}
 \caption{\\Summary of the maximal global synchronization error on line. The 95\% confidence interval is mean 16.81-17.49 {\textmu s} and standard deviation 3.49-3.97 in ATS, mean 11.09-11.74 {\textmu s} and standard deviation 2.94-3.4 in MACTS.}{\label{tab:2}}
 \setlength{\tabcolsep}{0.5mm}{
 \begin{tabular}{ccccc}
  \toprule
                  &    \multicolumn{3}{c}{\textbf{{The maximal global Synch-error}}}                                     & \textbf{Convergence}\\
                  & \tabincell{c}{Mean (\textmu s)}    &\tabincell{c}{STD} &\tabincell{c}{Maximal (\textmu s)}          & \textbf{time (Minute)}\\
  \midrule
        ATS	      &17.15	                             &3.71	            &31	                                       & 27-32  \\
        MACTS	  &11.42	                             &3.16	            &19	                                       & 6-12 \\
  \bottomrule
 \end{tabular}}
\end{table}

\section{Simulation and Convergence Time}
In order to further compare the performances of MACTS to ATS, a $60\times60$ grid network simulation (diameter of 118) was build. Where, the synchronization period is 30 second, and initial $H$ is 1, 2, 4, 6, separately. The transmission delay of adjacent nodes is modeled as a normal distribution with mean of 3.3 {\textmu s} and standard deviation of 0.0049. The relative clock frequency offset is $\pm40$ PPM, and the initializations of clock offset are generated in random value ($<500$ seconds). When initial $H$ is 1, then the MACTS is actually a single-hop communication time synchronization algorithm, i.e., ATS.

\begin{figure}[!htb]
\centering
\includegraphics[scale=0.6]{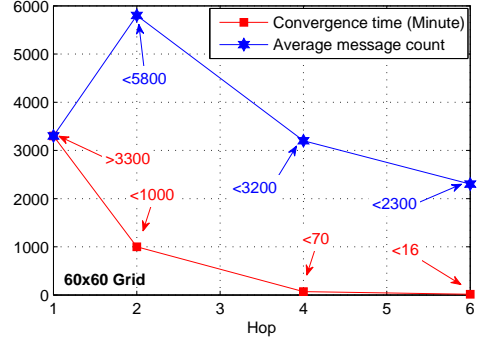}
\caption{Convergence time and message count of MACTS on $60\times60$ grid. The average message count is the communication cost at the moment of the synchronization convergence.}
\label{fig.5}
\end{figure}

Referring to the experimental results, the simulation’s convergence judgment condition was set to be a maximal global synchronization error of $<$20 {\textmu s}.
The convergence time of the simulation under different hop MACTS are shown in Fig. \ref{fig.5}. The convergence time of ATS is larger than 3300 minutes. While the 6-hop MACTS's convergence time is less than 16 minutes. Moreover, the average message count is the total amount of message transmission until the algorithm converges, and it is less than 2300 and more than 3300 in 6-hop MACTS and ATS, respectively. After about 30 minutes, $H$ is gradually decreased from 6 to 1 in 6-hop MACTS, meanwhile, the message count in MACTS is same as that in ATS.

\vspace{0.5cm}

It is clear that, too long convergence time is a serious problem for deploying average consensus-based time synchronization, as it will seriously limit the reliability and timeliness of the algorithm. Particularly, it is impossible to meet the timely synchronization requirements in the energy-efficiently large-scale WSN.

The convergence rate in MACTS is significantly improved with short-distance multi-hop communication. Moreover, the simulations on the random deployed topology networks also show that. Considering the large-scale WSN applications, it is cost-effective and reasonable to slightly increase the message complexity in exchange for a significant improvement in synchronization rate.

\section{Conclusions}
In this paper, we proposed a fast convergence average consensus-based time synchronization algorithm for large-scale WSN, i.e., MACTS. The multi-hop communication is used to generate a lots of virtual connections between non-adjacent nodes, and leads to the algebraic connectivity of networks increases greatly. Meanwhile, the MACTS multi-hop controller is developed to against the possible increases in message complexity and synchronization error. Moreover, the delay estimation is used to compensate the relative clock offset estimation and local synchronization error. Evaluating the MACTS by experiments and simulations, results indicate that the convergence rate and accuracy are significantly improved in MACTS. Specifically, the convergence rate in a 6-hop MACTS is hundreds of times of that in ATS.

\ifCLASSOPTIONcaptionsoff
  \newpage
\fi
{
\scriptsize
\bibliographystyle{ieeetr}
\bibliography{TXT_TII-19-0934}
}

\begin{IEEEbiography}[{\includegraphics[width=1in,height=1.25in,clip,keepaspectratio]{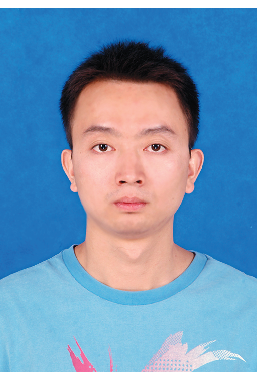}}]{Fanrong Shi}
 (S'18) received the B.E., M.E. and Ph.D. degrees in communication engineering, communication and information system, control science and engineering from Southwest University of Science and Technology (SWUST), Mianyang China, in 2009, 2012, 2019, respectively. Since 2012, he is currently an Assistant Professor in information and communication engineering with the Department of School of Information Engineering at SWUST. His research interests include time synchronization and location in wireless sensor networks, wireless measurement, and signal acquisition.
\end{IEEEbiography}

\begin{IEEEbiography}[{\includegraphics[width=1in,height=1.25in,clip,keepaspectratio]{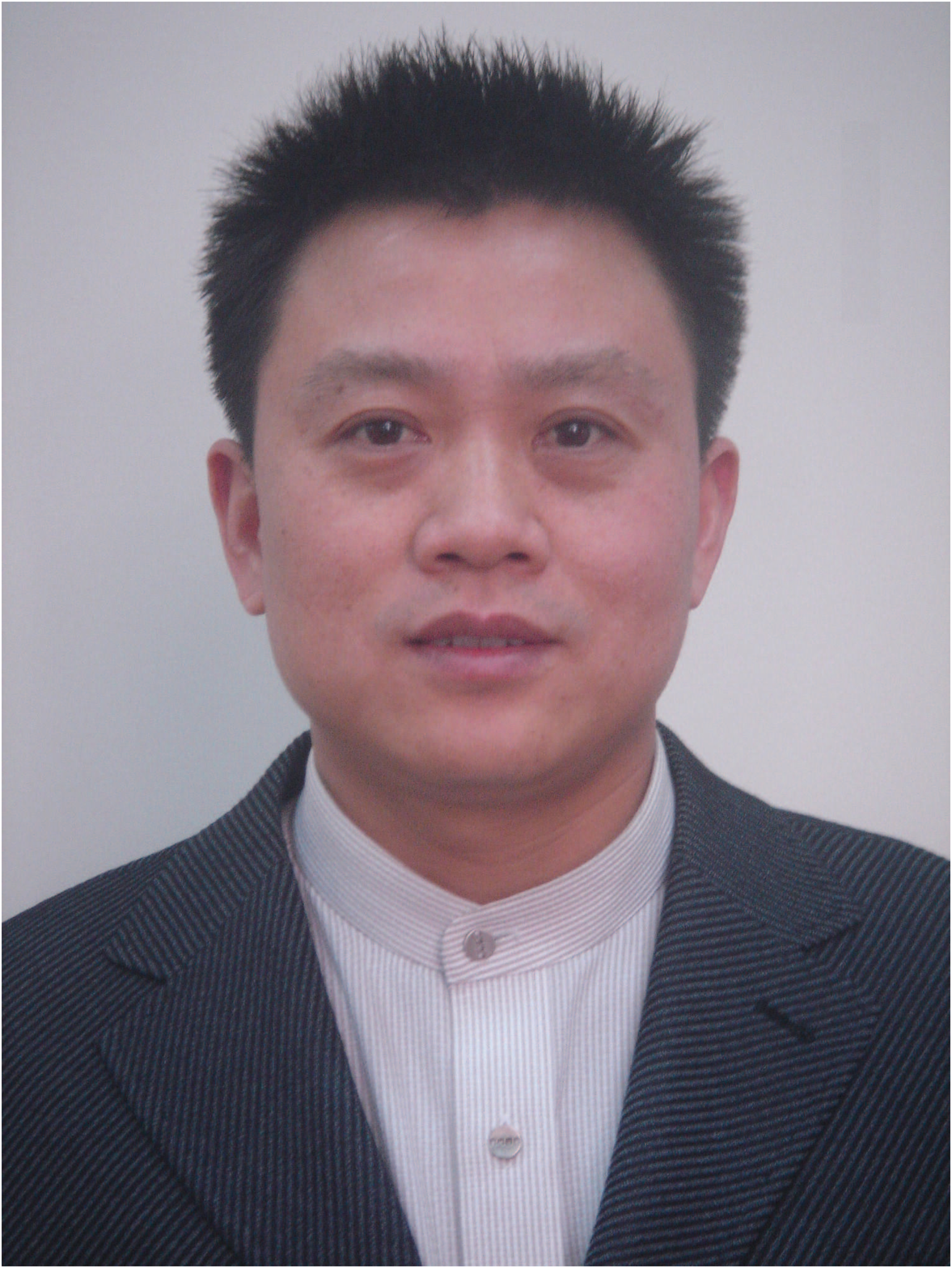}}]{Xianguo Tuo}
received the B.E., M.E. and Ph.D. degrees in nuclear geophysics, environmental radiation protection, applied nuclear technology in geophysics from the Chengdu University of Technology in Chengdu, China, in 1988, 1993, 2001 respectively. From 2006 to 2007, he worked as visiting scholar at the School of Bioscience at University of Nottingham, UK. Since 2001, he is Professor with College of Nuclear Technology and Automation Engineering, Chengdu University of Technology, China. Since 2012, he becomes Professor with the School of National Defense Science and Technology, Southwest University of Science and Technology in Mianyang, Sichuan, China. Currently, he is the Professor in artificial intelligence and robot technology and President of Sichuan University of  Science and Engineering, Zigong, Sichuan. Professor Tuo received The National Science Fund for Distinguished Young Scholars in 2011. Currently, his research interests are in detection of radiation, seismic exploration and specialized robots.
\end{IEEEbiography}

\begin{IEEEbiography}[{\includegraphics[width=1in,height=1.25in,clip,keepaspectratio]{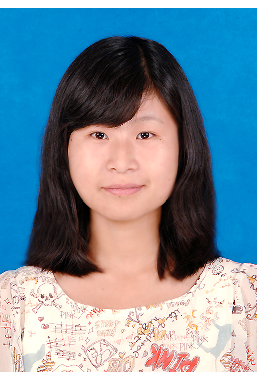}}]{Lili Ran}
received B.E. in Electric Engineering and M.E. in Rail Transit and Electrical Automation from the Southwest Jiaotong University. She is currently a member of the Robot Technology Used for Special Environment Key Laboratory of Sichuan Province, and a lecturer in control science and engineering with the Department of School of Information Engineering at Southwest University of Science and Technology. Her research interests include numerical simulation distributed sensing and system modelling.
\end{IEEEbiography}

\begin{IEEEbiography}[{\includegraphics[width=1in,height=1.25in,clip,keepaspectratio]{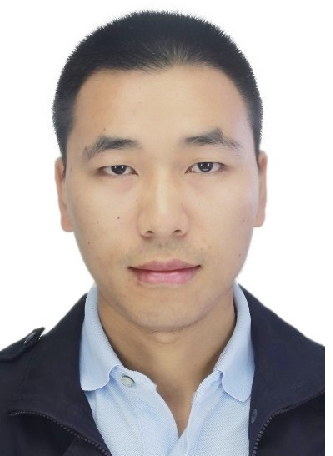}}]{Zhenwen Ren}
(S'19) received the M.S. degree in communication and information system at Southwest University of Science and Technology (SWUST), Mianyang, China. He is pursuing the Ph.D. degree in control science and engineering at Nanjing University of Science and Technology (NJUST), Nanjing, China. He is also currently with the Department of School of National Defence Science and Technology at SWUST. His research interests include image set classification, subspace clustering, and deep learning.
\end{IEEEbiography}

\begin{IEEEbiography}[{\includegraphics[width=1in,height=1.25in,clip,keepaspectratio]{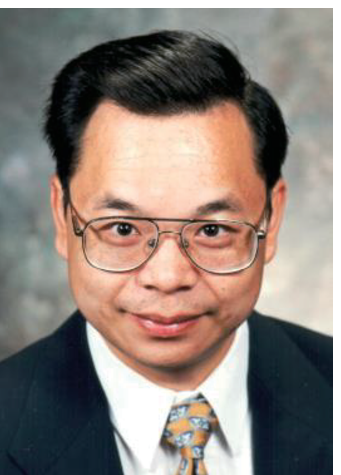}}]{Simon X. Yang}
(S'97, M'99, SM'08) received the B.Sc. degree in engineering physics from Beijing University, China in 1987, the first of two M.Sc.  degrees in biophysics from Chinese Academy of Sciences, Beijing, China in 1990, the second M.Sc. degree in electrical engineering from the University of Houston, USA in 1996, and the Ph.D. degree in electrical and computer engineering from the University of Alberta, Canada in 1999. Currently he is a Professor in systems and computer engineering and the Head of the Advanced Robotics and Intelligent Systems (ARIS) Laboratory at the University of Guelph in Canada. His research interests include intelligent systems, robotics, sensors and multi-sensor fusion, wireless sensor networks, control systems, and computational neuroscience. Prof. Yang serves as the Editor-in-Chief of International Journal of Robotics and Automation, and Associate Editor of IEEE Transactions on Cybernetics, and several other journals. He has involved in the organization of many conferences.
\end{IEEEbiography}

\end{document}